\documentstyle [12pt] {article}
\begin{document}
\rightmargin -2.75cm
\textwidth 18.cm
\textheight 23.0cm
\topmargin -0.5in
\baselineskip 16pt
\parskip 18pt
\parindent 30pt
\title{ \large \bf Algebraic approach to
quantum field theory on non-globally-hyperbolic spacetimes}
\author{Ulvi Yurtsever \\
{}~~~~~~~~~~~~ \\
Department of Physics \\
University of California \\
Santa Barbara, CA 93106 }
\date{October, 1992 \\
UCSBTH-92-43}
\pagestyle{empty}
\baselineskip 24pt
\maketitle
\vspace{.2in}

\baselineskip 12pt
\begin{abstract}
\noindent The mathematical formalism for linear quantum field theory on curved
spacetime depends in an essential way on the assumption of
global hyperbolicity. Physically,
what lie
at the foundation of any formalism for quantization in curved spacetime
are the canonical commutation
relations, imposed on the field operators evaluated at a global Cauchy surface.
In the algebraic formulation of linear quantum field theory, the
canonical commutation relations are restated in terms of
a well-defined symplectic
structure on the space of smooth solutions, and the local field algebra is
constructed as the Weyl algebra associated to this symplectic vector space.
When spacetime is not globally hyperbolic, e.g. when it contains naked
singularities or closed timelike curves, a global Cauchy surface does
not exist, and there is no obvious way to formulate the canonical
commutation relations, hence no obvious way to construct the field
algebra. In a paper submitted elsewhere, we
report on a generalization of the algebraic framework for quantum field
theory to arbitrary topological spaces which do not necessarily have
a spacetime metric defined on them at the outset. Taking this generalization
as a starting point, in this paper
we give a prescription for constructing the field
algebra of a (massless or massive) Klein-Gordon field
on an arbitrary background spacetime.
When spacetime is globally hyperbolic, the theory defined by our
construction
coincides with the ordinary Klein-Gordon field theory on a
globally hyperbolic background.
We explore some basic features of our
generalized Klein-Gordon theory on arbitrary spacetimes, and study its
specific properties on simple examples of
non-globally-hyperbolic backgrounds that contain closed timelike curves
or naked singularities.
\vspace{0.5cm}

\end{abstract}

\newpage
\pagestyle{empty}
\baselineskip 16pt
\parskip 16pt
\pagestyle{plain}
\pagenumbering{arabic}
{}~~~~~~

{\bf \noindent 1. The algebraic approach to quantum field theory}

The quantum theory of a linear Klein-Gordon field on a globally hyperbolic
spacetime $(M,g)$ can be built entirely on the basis of
the well known canonical
commutation relations (CCR)
\begin{eqnarray}
[{\phi} (x) , {\phi} (x^{\prime})] & = & 0 \; , \nonumber \\ ~
[{\phi} (x) , \dot{\phi} (x^{\prime})] & = & i \delta (x, x^{\prime})
\; , \; \; \; \; \mbox{for} \; x, \; x^{\prime} \in \Sigma \; ,
\end{eqnarray}
where $\phi (x)$ denotes the field operator ``evaluated" at the point
$x$, $\Sigma$ is a Cauchy surface for spacetime, and $\dot{\phi}$
denotes the normal time derivative of the field evaluated on $\Sigma$.
The CCR provide
well-posed initial data on $\Sigma$ for the antisymmetric two-point function
$[ {\phi} (x) , {\phi} (x^{\prime}) ]$, which can then be evolved
via the Klein-Gordon field equation
\begin{equation}
( {\nabla}^{a}{\nabla}_{a} - m^{2} ) \phi = 0
\end{equation}
to yield the complete solution. It is also a consequence of the field
equation (2) that all commutators on $\Sigma$ involving polynomial
functions of the field $\phi$ and its normal
time derivative (``canonical momentum") $\pi = \dot{\phi}$ can be
determined solely from Eqs.\,(1).

Everyone knows, of course, that the field operator $\phi$ is ill-defined
when evaluated at a point. To construct a mathematically well-defined
quantum field theory, we need to work with the smeared field operators
\begin{equation}
\phi (f) \equiv \int_{M} f(x) \phi (x) \, d \tau \; ,
\end{equation}
where $f$ is a real-valued
$C^{\infty}$ test function of compact support on $M$, and
$d \tau$ denotes the spacetime four-volume form of $(M,g)$. To express the
CCR [Eqs.\,(1)] in terms of these smeared field operators, we introduce the
vector space, $S_{0}(M,g)$, which consists of
the $C^{\infty}$ solutions of Eq.\,(2)
that are compact-supported on $\Sigma$ (and hence are compact-supported
on any Cauchy surface). On $S_{0}(M,g)$, there exists a natural,
antisymmetric bilinear form (the Klein-Gordon inner product)
\begin{equation}
\sigma (F,G) \equiv \int_{\Sigma} (F {\nabla}^{a} G -
G {\nabla}^{a} F ) \, d {\Sigma}_{a} \; \; , \; \; \; \;
F, \; G \in S_{0}(M) \;
,
\end{equation}
where $d{\Sigma}_{a}$ is the three-volume form of $\Sigma$. The integral
in Eq.\,(4) exists because $F$ and $G$ are
compact supported on $\Sigma$, and it
is independent of the choice of Cauchy surface $\Sigma$ because $F$ and $G$
satisfy the wave equation
(2). The Klein-Gordon form $\sigma$ makes $S_{0}(M)$ into a symplectic vector
space. To connect this symplectic structure with the CCR,
we let $C_{0}^{\infty}(M)$ denote the space of
smooth test functions of compact support on $M$. It
follows [1] from the global hyperbolicity of
$(M,g)$ that for each $f \in C_{0}^{\infty}(M)$
there exists a unique solution,
${\cal E}^{+} f$, of the inhomogeneous wave equation
\begin{equation}
({\nabla}^{a}{\nabla}_{a} - m^{2} ) \phi = f \;
\end{equation}
that vanishes outside $J^{-}( \mbox{support} (f) )$, called the ``advanced"
solution with source $f$, and a unique
solution, ${\cal E}^{-} f$, that
vanishes outside $J^{+} ( \mbox{support} (f) )$, called the ``retarded"
solution. Moreover, the advanced and retarded
solutions ${\cal E}^{\pm} f$ are compact
supported on Cauchy surfaces since $f$ has compact support in $M$.
Hence, the difference ``advanced $-$ retarded solution with source
$f$" defines a linear map
\begin{equation}
{\cal E} \equiv {\cal E}^{+} - {\cal E}^{-} \; : \;
C_{0}^{\infty} (M) \longrightarrow S_{0}(M,g)
\end{equation}
which associates to each $f \in C_{0}^{\infty}$ the solution
${\cal E}^{+}f - {\cal E}^{-}f$ of Eq.\,(2)
lying in $S_{0}(M)$. It is not difficult to show (i) that ${\cal E}$ is
onto, i.e. for each $F \in S_{0}(M)$ there exists a $f \in
C_{0}^{\infty} (M)$ for which $F = {\cal E} f$,
(ii) that ${\cal E} f = 0$ if and only if there exists a $g
\in C_{0}^{\infty} (M)$ such that $f=(
{\nabla}^{a}{\nabla}_{a}  - m^{2} ) g$, and (iii) that
the identity
\begin{equation}
\int_{M} f(x) G(x) \, d \tau = \sigma ( {\cal E} f, G)
\end{equation}
holds for all $f \in C_{0}^{\infty}(M)$ and for all $G \in S_{0}(M)$.
Finally, combining Eqs.\,(3), (4) and (7) with Eqs.\,(1), we obtain the
identity
\begin{equation}
[ \phi (f) , \phi (g) ] = i \Delta (f,g) \; , \; \; \; \; f, \; g \in
C_{0}^{\infty} (M) \; ,
\end{equation}
where $\Delta$ denotes the two-point distribution [bi-solution of
Eq.\,(2)]
\begin{equation}
\Delta (f,g) \equiv \sigma ( {\cal E} f , {\cal E} g ) \;, \; \; \; \;
f, \; g \in C_{0}^{\infty} (M) \;  .
\end{equation}
Equation (8) is the desired expression of the CCR in terms of the
smeared field operators $\phi (f)$.

Quantization of the Klein-Gordon field theory will now be complete once we
construct the
Hilbert space on which $\phi (f)$ act as operators. This can be done
directly, by building a Fock space representation of the CCR [Eq.\,(8)],
when a specific ``vacuum" state in Hilbert space is singled out by
the physics of $(M,g)$. For example, in flat, Minkowski spacetime, the
Poincare-invariant Minkowski vacuum can be used to build the standard
Fock-space representation of the CCR. However, in a general, globally
hyperbolic spacetime, no such preferred vacuum state
exists. Furthermore, Fock representations of the CCR built from
different vacuum states are in general unitarily inequivalent. The only
consistent, general way of implementing the CCR into a well-defined
mathematical formalism is the algebraic approach. We shall now briefly
review the algebraic approach, but only those
basics which will be referred to later
in the paper. Ref.\,[2] contains a thorough discussion and ample references;
for a more peaceful introduction see [3].

Fundamental to the
algebraic approach is the notion of the ``Weyl algebra." Given any symplectic
vector space $(S, \sigma )$ over ${\bf R}$,
the {\it pre-Weyl algebra}
associated to $(S, \sigma )$ is defined as the $\ast$-algebra
${\cal W}^{\circ} (S , \sigma )$
over the complex field ${\bf C}$ consisting of finite sums of the form
\begin{equation}
\sum_{i} {\alpha}_{i} W(F_{i}) \; , \; \; \; \; {\alpha}_{i} \in {\bf
C}, \; F_{i} \in S \; ,
\end{equation}
where the generators $W(F)$ satisfy the relations
\begin{eqnarray}
W(F)W(G) & = & e^{ - \frac{i}{2} \sigma (F,G) } \, W(F+G) \nonumber \\
         & = & e^{ - i \sigma (F,G) } \, W(G)W(F) \;
\end{eqnarray}
from which the product of two general elements in the form of Eq.\,(10) can
be obtained using linearity and the distributive law. The $\ast$ operation
of ${\cal W}^{\circ} (S, \sigma )$ is simply
\begin{equation}
\left[ \sum_{i} {\alpha}_{i} W(F_{i}) \right]^{ \ast } \equiv
\sum_{i} \overline{{\alpha}_{i}} \: W( - F_{i} ) \; ,
\end{equation}
and the identity element $\bf 1$ is $W(0)$. It is a consequence of the
nondegeneracy of the symplectic form
$\sigma$ on $S$ that the pre-Weyl algebra
${\cal W}^{\circ} (S, \sigma )$ is simple, i.e.
it does not contain any proper two-sided ideals [see [4] for a
short proof of this]. As a result, there exists a unique $C^{\ast}$
norm on ${\cal W}^{\circ} (S, \sigma ) $;
the {\it Weyl algebra}, ${\cal W} (S, \sigma )$,
is defined as the $C^{\ast}$ algebra obtained by completing
${\cal W}^{\circ} (S, \sigma )$ under this norm.

Consider now the Weyl algebra ${\cal A}$ corresponding to the
symplectic space of solutions, $(S_{0}, \sigma )$, of the
Klein-Gordon equation (2), where $\sigma$ is defined by
Eq.\,(4). Using Eq.\,(7), we can formally rewrite
the smeared field operators $\phi (f)$ in the form
\begin{equation}
\phi (f) = \int_{M} f(x) \phi (x) \, d \tau = \sigma ( {\cal E} f,\phi )
\; ,
\end{equation}
which shows that $\phi (f)$ depends only on the solution ${\cal E} f$
in $S_{0} (M)$. We can therefore consider the field operators as
being smeared with solutions $F \in S_{0} (M)$ instead of with test functions
$f \in C_{0}^{\infty} (M)$; hence $\phi (F) , \; F \in S_{0}$, denotes
the operator $\phi (f)$, where $f \in C_{0}^{\infty} (M)$ is any test
function that satisfies ${\cal E} f = F$. The generators $W(F)$ of the
Weyl algebra ${\cal A}$ then correspond, formally, to the exponentiated
fields
\begin{equation}
W(F) = e^{i \phi (F) } = e^{i \phi (f) } \; , \; \mbox{where} \;
F \in S_{0} (M) , \; f \in C_{0}^{\infty} (M), \; {\cal E} f = F \;
\end{equation}
as the reader can verify by comparing the commutation relations (8) with
the corresponding relations (11) for $W(F)$. [Note that the CCR in the
form of Eq.\,(8) cannot be represented by an algebra of
bounded operators, so
in order to to build a
$C^{\ast}$ operator algebra with a well defined norm
it is necessary to use the exponentiated version of the CCR
given by Eq.\,(11).] The $C^{\ast}$ algebra
${\cal A} = {\cal W} (S_{0}, \sigma )$ defines
the local algebra of field operators for the quantized Klein-Gordon
field on $(M,g)$. [The term ``local" refers to the existence of the
preferred set of closed subalgebras, $\{ {\cal A} (U) \}$, in ${\cal A}$,
where for each open subset $U \subset M$,
${\cal A} (U)$ denotes (the closure
of) the subalgebra
generated by those elements $W( {\cal E} f )$ for which $f$
is supported in $U$. This extra structure is of key importance to
the generalization of the algebraic approach which we will discuss
in the next section.] We will use the more descriptive notation ${\cal
A}^{(KG, m)}$ for the Klein-Gordon algebra
when it is necessary to indicate the
explicit field theory it refers to.

All information one might wish to have about
the quantized Klein-Gordon field
on $(M,g)$ is contained in the local algebra ${\cal A}^{(KG,m)}$. Thus, for
example, quantum states of the Klein-Gordon field are
nothing but (complex) linear functionals $\omega$ on $\cal A$ that
satisfy the positivity and normalization conditions
\begin{equation}
\omega ( x^{\ast} x ) \geq 0 \; \; \; \; \;
\forall x \in {\cal A} \; , \; \; \; \; \omega ({\bf 1}) = 1 \; .
\end{equation}
The $n$-point functions associated to a state $\omega$ can be defined
directly in terms of the algebraic structure of $\cal A$. For example,
the two-point function, when it exists, is defined as the distribution
[bi-solution of Eq.\,(2)]
\begin{equation}
{\lambda}_{2} (f,g) = - \left. \frac{{\partial}^{2}}{\partial s \partial t}
\right|_{s=t=0} \left [ \, \omega [W(s \, {\cal E} f + t \, {\cal E} g )]
\: e^{ - \frac{i}{2} st \: \sigma ( {\cal E} f , {\cal E} g ) } \, \right]
\end{equation}
acting on test functions $f, \; g \in C_{0}^{\infty} (M)$.
States also provide the connection with Hilbert-space quantization through
the so-called ``GNS construction" ([4]$-$[6]), which associates to each state
$\omega$ a Hilbert space, $[H_{\omega}, <,>]$, a ``vacuum" vector
$\Omega \in H_{\omega}$, and a representation
${\rho}_{\omega}$ of ${\cal A}$ by bounded operators on $H_{\omega}$
such that $\omega (x) = < \Omega , {\rho}_{\omega} (x) \Omega > \;
\forall x \in {\cal A}$.
When the two-point function ${\lambda}_{2} (f,g)$
of ${\omega}$ exists [Eq.\,(16)], a one-particle
Hilbert space ${\cal H}_{\omega}$ can be built from the space
of solutions $S_{0}$ using the real
part of ${\lambda}_{2}$ as an inner product;
the GNS construction $( H_{\omega} ,
{\rho}_{\omega} )$ is then identical to the standard Fock-space
construction based on the one-particle Hilbert space ${\cal H}_{\omega}$
(see Sect.\,3.\,2 of [2] for details).

The compact, unified way in which the notion of a ``quantum state" is
formalized is the main advantage of the algebraic approach, and this
advantage is present independently of whether the algebra $\cal A$ has
unitarily inequivalent representations (as a concrete algebra of
operators on Hilbert space). [A Weyl algebra ${\cal W} (S, \sigma )$
has inequivalent representations whenever $S$ is infinite dimensional.]
The algebraic viewpoint is in fact indispensable for discussing quantum
field theory on any background other than flat, Minkowski spacetime.

{\bf \noindent 2. A generalization of the algebraic approach}

The quantization formalism outlined above relies in a crucial way on the
existence of
a global Cauchy surface in spacetime. Globally hyperbolic spacetimes,
which are the ones that admit such surfaces, are adequate for discussing
most physically interesting problems in General Relativity. Moreover, there
exists a rather large collection of arguments,
organized around the well-known ``cosmic
censorship hypothesis," which suggest that under generic conditions global
hyperbolicity may in fact be enforced as a
consequence of other, more fundamental laws of physics.
While still supported by the
standard evidence for cosmic censorship, the argument for global
hyperbolicity has been challenged increasingly in recent years, and
understanding the physics of spacetimes containing
closed timelike curves or naked
singularities no longer appears to be a question of purely academic interest.
Clearly, it is important to remove the global hyperbolicity restriction
from the algebraic formulation of quantum field theory.

How can we construct the algebra of the quantum Klein-Gordon field when
spacetime lacks a global Cauchy surface $\Sigma$, hence lacks a well-defined
symplectic space of solutions $(S_{0} , \sigma )$? Equivalently, how can we
formulate quantum field theory in the absence of CCR? The familiar
path-integral formalism suggests one way of approaching such a
generalization. However, we desire more mathematical
clarity and consistency, at least for linear (non-interacting) field theory,
than the path-integral approach can provide.
For us, the answer lies in widening
the notion of ``quantum field" so that the construction of the field
operators is not tied to the (canonical) quantization of a Hamiltonian
system for classical fields. (There is no well defined dynamics for
classical fields on a
non-globally-hyperbolic spacetime.) In a separate paper submitted
elsewhere [7], we report on an extension of the algebraic notion of quantum
field to general background spaces. The only structure these backgrounds
posses is that of a topological space, i.e. they
may not even have a spacetime metric defined on
them at the outset. The primary motivation for such a drastic
generalization is to understand the ultimate origin of spacetime
topology, and the motivations for and the consequences of this approach
are discussed in detail in [7]. Here we will only summarize the relevant
aspects of this generalized algebraic notion of quantum field theory as
it applies to the specific problem we are addressing.

Let $X$ be a topological space. A ``quantum field theory" on $X$ consists
of an abstract $C^{\ast}$ algebra ${\cal A}$ (with identity element $\bf
1$), and a map (which we will
also denote by ${\cal A}$) that associates to each open subset $U$ in
$X$ a closed subalgebra ${\cal A} (U) \subset {\cal A}$ such that the
following two conditions hold:

\noindent (QF1): For every open subset $U \subset X$ $\; {\cal A} (U)$
is a central $C^{\ast}$ algebra, and ${\cal A}
(\{ \}) = \bf C \cdot \bf 1$, ${\cal A} (X) = {\cal A}$.

\noindent (QF2): For any collection $\{ W_{\alpha} \}$ of open subsets,
\begin{equation}
{\cal A} ( \bigcup_{\alpha} W_{\alpha} ) = \overline{ <
\bigcup_{\alpha} {\cal A} (W_{\alpha}) > } \; .
\end{equation}
Here $\{ \}$ denotes the empty set, $\bf C \cdot \bf 1$
is the $C^{\ast}$ algebra (isomorphic to the algebra $\bf C$ of complex
numbers) generated by $\bf 1$,
a {\it central} algebra ${\cal B}$ is one with
the property that its center, \[ Z({\cal B}) \equiv
\{ x \in {\cal B} \, | \, xy=yx \; \forall y
\in {\cal B} \} \; , \] is equal to
$\bf C \cdot \bf 1 \, $ [$Z({\cal B}) \cong {\bf C}$],
$\; <S>$ denotes the
subalgebra generated by a subset $S \subset
{\cal A}$, and overbar denotes closure.
We will call the theory ${\cal A}$ ``nondegenerate" if ${\cal A}(U)$ is
strictly bigger than ${\bf C}$ for every nonempty open subset $U \subset
X$. Note that, in standard quantum
field theory, property QF2 does
not in general hold for the local algebra of all ``observables," but it
does hold when ${\cal A}$ consists only of (exponentiated)
smeared field operators.

Next we incorporate the notion of ``locality" into our generalized field
theory. For this, let for each point $p \in X \, $ $C(p)$ denote the set
\begin{eqnarray}
C(p) & \equiv & \{ q \in X | \not{\exists} \; \mbox{open sets} \; U , \; V \;
\mbox{such that} \nonumber \\
&  & p \in U , \; q \in V  , \; \mbox{and} \;
[{\cal A} (U), {\cal A} (V)] = 0  \} \; ,
\end{eqnarray}
where for $A, \; B \subset {\cal A} $, $\, [A,B]$ denotes the commutator
subalgebra generated by elements of the form $\{ ab-ba \, | \, a \in A,
\; b \in B \}$. The set $C(p)$ consists of those points $q \in X$ that
can ``causally communicate" with $p$
through fields in ${\cal A}$. [We restrict ourselves throughout to bosonic
fields; hence our use of the commutator $[ \; , \; ]$. It is straightforward to
formulate a fermionic version of our discussion by replacing commutators
with anti-commutators. But note that, in the fermionic case, the generators
of the field algebra are the smeared field operators themselves
(which are already bounded) instead of the exponentiated fields (14).]
Some immediately obvious
properties of $C(p)$ are: $q \in C(p) \; \mbox{iff} \; p \in C(q)$,
$C(p)$ is a closed subset of $X$, and,
when ${\cal A}$ is nondegenerate, $p \in C(p) \; \; \forall p \in X$
(this last result follows from QF1). A continuous curve $\gamma :
{\bf R} \longrightarrow X$ is called a ``connector" if
for every $t_{0} \in {\bf R}$ there exists an $\epsilon > 0$ such that
$\gamma (t) \in C[ \gamma (s)] $ for all $t$, $s$ in the interval
$(t_{0} - \epsilon \, , \, t_{0}+ \epsilon )$.
(Thus defined a connector is analogous to a causal curve in spacetime.)
The notion of locality for a quantum field theory ${\cal
A}$ on $X$ is now defined in terms of the topological properties of the
sets $C(p)$. Thus, we will say that ${\cal A}$ is ``weakly local" if the
following two conditions are satisfied:

\noindent (L): There exists an open
neighborhood $V$ around every point $p \in X$
such that for every open neighborhood
$U$ of $p$ contained in $V$ the set $U \cap [C(p) \backslash \{ p \}]$ is
disconnected (here $\backslash$ denotes set difference).

\noindent (WL): For all $p \in X$ $\, C(p)$ is connected.

\noindent The theory $\cal A$ is ``strongly local" if it satisfies
condition L and the following stronger version of WL:

\noindent (SL): For every $p$, $q \in X$ such that $q \in C(p)$ there
exists a connector $\gamma$ joining $p$ and $q$; in particular, the set
$C(p)$ is arcwise connected for all $p \in X$.

\noindent It is easy to see that if condition
L is satisfied ${\cal A}$ must be
nondegenerate.

In physical terms, locality provides
for the existence of dynamics, the
``finite speed of propagation" of causal signals.
Accordingly, condition WL (or SL) guarantees that
causal influences propagate from $p$
continuously, and condition L guarantees that signals that communicate
with $p$ propagate with ``finite speed," and that they connect $p$ to disjoint
components of $C(p) \backslash \{ p \}$ (which is necessary if dynamics
at $p$ is to be determined not only by local evolution equations but also
by boundary conditions).

The next notion we will need is that of ``isomorphism" between quantum field
theories. Let $X$ be a topological space, and let ${\cal A}_{1}$ and ${\cal
A}_{2}$ be field theories on $X$. Then, ${\cal A}_{1}$ and ${\cal A}_{2}$
are said to be isomorphic, denoted
$({\cal A}_{1}, X) \cong ({\cal A}_{2},X)$,
if there is a (isometric) $C^{\ast}$ isomorphism
$\Psi : {\cal A}_{1} \longrightarrow {\cal A}_{2}$ such that for every
open subset $U \subset X$, \[ {\cal A}_{2} (U) = \Psi [ {\cal A}_{1} (U)
] \; . \] Note that, if ${\cal A}$ is an ordinary
(e.g. Klein-Gordon) field theory on a globally hyperbolic spacetime
$(M,g)$, $\, h : M \longrightarrow M$ is a diffeomorphism of $M$, and
if we define a new theory $h^{\ast} {\cal A}$ on $M$ by (for
all open $U \subset M$) \[
(h^{\ast} {\cal A} ) \: (U) \equiv \left. {\cal A}^{(KG,m)} \right|_{(M,
h^{\ast}g)}  [h^{-1} (U)] \]
[where the field theory that
appears on the right-hand
side is the standard Klein-Gordon theory corresponding to the spacetime $(M,
h^{\ast}g)$], then the theories ${\cal A}$ and
$h^{\ast} {\cal A}$ are isomorphic over $M$.
Hence our notion of isomorphism is a
generalization of the usual diffeomorphism invariance in curved-spacetime
field theory.

We will also need the following result, whose proof can be found (along
with a more detailed discussion of the material of this section) in [7]:

{\noindent \it Theorem}: Let $X$ be a locally compact topological space,
and $\cal A$ a weakly local quantum field theory on $X$. If $U$, $V
\subset X$ are open sets such that ${\cal A} (U)$ and ${\cal A} (V)$ {\it
fail} to commute, then there exists a point $p \in U$ and a point $q \in
V$ such that $q \in C(p)$.

The generalized Klein-Gordon field theory on a non-globally-hyperbolic
spacetime is a field theory in the extended sense we have described in
this section. We now turn to the explicit construction of this theory.

{\bf \noindent 3. Construction of the Klein-Gordon field algebra on
an arbitrary spacetime}

Let $(M,g)$ be a spacetime (not necessarily globally hyperbolic).
We define a subset $U \subset (M,g)$ to be ``locally causal" if $U$ is
open, connected, and the
spacetime $(U, g|_{U})$ is globally hyperbolic.
[Note that this notion is distinct from the notion of a ``globally
hyperbolic subset" in $(M,g)$, defined, in the standard way, as a subset
$U$ such that (i) strong causality holds
on $U$, and (ii) $\forall \; p$ and $q \in
U$ the set $J^{-}(q) \cap J^{+}(p)$ is contained entirely in $U$ and is
compact ([8], Chapter 6). In general, a locally causal subset
does not satisfy any of these properties; even when the ambient
spacetime $(M,g)$ is globally hyperbolic, only (i) and the second half of
(ii) hold generally for locally causal $U \subset M$.]
Clearly, every spacetime admits an open covering by locally causal subsets.
We assume that a space
of global solutions, $S(M)$, of Eq.\,(2) has been fixed, through physical
considerations, as the configuration space of classical fields on
$(M,g)$. Usually, $S(M)$ will simply consist of the space of all $C^{\infty}$
global solutions of the Klein-Gordon equation (2); however, in some
cases it may be necessary to include less smooth ($C^{k}$ or
distributional) solutions in $S(M)$. The resulting Klein-Gordon
field theory on $M$
will depend in an essential way on this initial choice of $S(M)$.

Let $T$ denote the real vector space consisting of finite sums of the
form
\begin{equation}
\sum_{i=1}^{n} c_{i} \, (F_{i}, U_{i}) \; ,
\end{equation}
where each $U_{i}$ is a locally causal open subset, $c_{i} \in {\bf R}$,
and each $F_{i}$ is
an element in $S_{0}(U_{i}) = S_{0} (U_{i}, g|_{U_{i}})$, the space of
all $C^{\infty}$ solutions of Eq.\,(2) defined and
compact supported on Cauchy surfaces in $(U_{i}, g|_{U_{i}})$. In $T$ we
make the identifications
\begin{eqnarray}
c \, (F, U) & = & (cF, U) \;, \; \; \;
(F_{1}, U)+(F_{2}, U) = (F_{1} + F_{2} , U) \; ,
\; \; \; (0, U) = 0 \; \nonumber \\
& \forall & U \; \mbox{locally causal}, \;
F, \; F_{1}, \; F_{2} \in S_{0}(U), \; c \in {\bf R}
\end{eqnarray}
and only these [thus, in more fancy terms, $T$ is the direct sum
$\bigoplus_{U} S_{0} (U) $ where $U$ ranges over all locally causal
subsets of $(M,g)$]. Let $S^{\ast}(M)$ denote the algebraic dual of
$S(M)$, i.e. the space of all linear functionals on $S(M)$.
Each element $x \in T$ defines a linear functional $\alpha_{x} \in
S^{\ast}(M)$ via the relation, $\forall G \in S(M)$,
\begin{equation}
\alpha_{x} (G) \equiv \sigma_{U} (F, G) = \sigma_{U} (F, G|_{U})
\; \; \; \; \;
\mbox{when} \; x = (F,U) \; ,
\end{equation}
where $\sigma_{U}$ denotes the Klein-Gordon inner product of the globally
hyperbolic spacetime $(U, g|_{U})$, and the action
of ${\alpha}_{x}$ on $S(M)$ is obtained by linear extension
when $x$ is an element of $T$ in the general form Eq.\,(19). If $U$ and
$V$ are locally causal subsets with $U \subset V$, then there exists a
natural imbedding
$i_{UV} : S_{0} (U) \hookrightarrow S_{0} (V) $ which satisfies
\begin{equation}
\alpha_{(i_{UV} F , \: V)} = \alpha_{(F,U)} \; \; \; \; \forall F \in S_{0}
(U) \; .
\end{equation}
This imbedding $i_{UV}$ is constructed as follows:
For any $F \in S_{0} (U)$, find a $f \in C_{0}^{\infty} (U)$
such that $F = {\cal E}_{U} f$, where ${\cal E}_{U} f$ denotes the
``advanced minus retarded solution with source $f$" (see Sect.\,1)
corresponding to the globally hyperbolic spacetime $(U, g|_{U})$. Then
simply put $\, i_{UV} F \equiv {\cal E}_{V} f$. This construction of
$i_{UV}F$ is well defined because ${\cal E}_{U} f = 0$ iff $f
= ({\nabla}^{a}{\nabla}_{a}
- m^{2}) g$ for some $g \in C_{0}^{\infty}(U)$, which implies
that ${\cal E}_{V} f = 0$. To prove that Eq.\,(22) holds, simply
observe that, using Eq.\,(7), we can write $\sigma_{U} ( {\cal E}_{U} f,
H)$ as $\int_{M} f H \, d \tau$ for all $f \in C_{0}^{\infty}(U)$ and
$H \in S(M)$.

It is now possible to define an antisymmetric bilinear form $\Omega$
on the space $T$ by putting
\begin{equation}
\Omega \, [(F,U), (G,V)] \equiv \left\{
\begin{array}{ll}
{\sigma}_{V} (i_{UV}F, \, G) \; , & \mbox{if $U \subset V$} \\
{\sigma}_{U} (F, \, i_{VU}G) \; , & \mbox{if $V \subset U$} \\
0 \; , & \mbox{otherwise ,}
\end{array}
\right.
\end{equation}
and extending $\Omega$ linearly to all
elements of $T$ given in the general form
Eq.\,(19). That $\Omega$ thus defined is a nondegenerate bilinear form
on $T$ is easily verified. Hence $(T, \Omega )$ is a
symplectic vector space.
The Weyl algebra associated to $(T, \Omega )$
might appear, at this point, to be a natural
candidate for the generalized Klein-Gordon algebra of $(M,g)$.
There are two obvious problems with this,
however: First, the space $(T,
\Omega )$ is unreasonably big, due clearly to those
elements which, from a
physical viewpoint, represent the same ``solution," but which are
contained as distinct vectors in $T$. Second, $(T, \Omega )$ does not
reduce to the standard symplectic space $[S_{0}(M) , \sigma ]$
(Sect.\,1) when $(M,g)$ is globally hyperbolic. We need to somehow
shrink $T$ down to the ``right" size so that every element it contains
represents a distinct local solution.
Guidance for doing this is
provided by the ``principle
of self consistency," which says, in essence, that the
only physically allowed
local solutions of the field equations
are those which
admit a smooth global extension. (See [9] and [10] for a detailed
discussion of this viewpoint in the context of spacetimes with closed
timelike curves.) Accordingly, we introduce the subspace $N \subset T$
given by
\begin{equation}
N = \{ x \in T \, | \, {\alpha}_{x} \equiv 0 \} \; .
\end{equation}
It is not difficult to verify that $N$ is a symplectic subspace of $T$,
i.e. the symplectic form $\Omega$ is nondegenerate when restricted to $N$.
The key idea behind the definition Eq.\,(24) of $N$
is that elements of $T$ which differ by a vector in $N$
are physically equivalent, because their difference cannot be detected using
the ``local" Klein-Gordon inner products with global solutions. It
would now appear that the most
natural way to proceed forward
from this point is to construct the quotient of
the Weyl algebra of $(T, \Omega )$ by the two-sided ideal generated by
elements $\{ W(n) \, | \, n \in N \}$,
i.e. to ``mod-out" the unphysical degrees
of freedom $W(n)$ from the field algebra. However, as we pointed out in
Sect.\,1, the Weyl algebra of any symplectic space is simple, i.e. does
not admit any nontrivial quotients. In particular, the two-sided ideal
generated by $\{ W(n) \, | \, n \in N \}$
in the Weyl algebra of $(T, \Omega )$ is
the entire algebra itself. This is a serious technical problem, and the
only way to circumvent it is to perform the moding-out construction at
the level of the vector space $(T, \Omega )$ instead of directly on the
Weyl algebra.

So let $Q$ denote the quotient vector space $T / N$. There is a natural
symplectic form $\sigma$ on this quotient,
namely the ``orthogonal projection" of the symplectic
form $\Omega$. For simplicity, assume that $T$ admits an orthogonal
decomposition
\begin{equation}
T = N \oplus N^{\perp} \; ,
\end{equation}
where $N^{\perp} \equiv \{ x \in T \, | \, \Omega (x,n) = 0 \; \;
\forall n \in N \}$. [Note that the decomposition (25) is necessarily a
direct sum (provided it exists) since $\Omega$ is nondegenerate on both
$T$ and $N$.] Then every $x \in T$ has a unique orthogonal projection
$x_{\perp} \in N^{\perp}$ such that $( x - x_{\perp} ) \in N$. Also, for
each equivalence class $[x] \in T/N$ and $x \in [x]$,
the projection $x_{\perp}$ depends
only on the equivalence class $[x]$ and not on the particular
representative $x$, i.e. for each $[x] \in Q$ we have a unique
projection $x_{\perp} \in N^{\perp}$. The
symplectic form $\sigma$ on $T/N$ is now defined simply by
\begin{equation}
\sigma ([x], [y]) \equiv \Omega (x_{\perp} , y_{\perp} ) \; .
\end{equation}
For infinite-dimensional subspaces $N \subset T$
the decomposition (25) does not always exist, however, and a more
complicated argument is necessary to construct $\sigma$ in the general
case. [When $N$ is finite dimensional, the decomposition (25) clearly
exists and can be constructed by purely algebraic means.]
We will describe this construction at the very end of this section;
for the rest of the section we will assume the construction as given.
In fact, throughout the rest of our discussion in Sect.\,3
the definition (26) is quite
sufficient since for the only concrete example in this section (that of a
globally hyperbolic spacetime) the decomposition (25) does exist.

We are now ready to construct the Klein-Gordon field algebra on $(M,g)$.
Put
\begin{equation}
{\cal A} \equiv {\cal W} (Q, \sigma ) \; ,
\end{equation}
the Weyl algebra corresponding to $(Q, \sigma )$, and let
for every open subset $V \subset M$
\begin{eqnarray}
T( V ) & \equiv & \{ x \in T \, | \, x = \sum_{i} (F_{i}, U_{i}) \;,
\; \mbox{where each} \; U_{i} \subset V \, , \;
F_{i} \in S_{0}(U_{i}) \} \; , \nonumber \\
Q(V) & \equiv & T(V) / N \; , \nonumber \\
{\cal A} (V) & \equiv & {\cal W} [Q(V), \sigma ] \; .
\end{eqnarray}
It is easy to verify the conditions QF1 and QF2 of Sect.\,2 for the
theory ${\cal A}$ defined by Eqs.\,(28). This is our generalized
Klein-Gordon field theory on the spacetime $(M,g)$, and we will continue
to denote it by ${\cal A}^{(KG,m)}$.

It may be of interest to note the following heuristic interpretation
for the algebra ${\cal A}$: Consider the symbols
$W([x])$, where $[x] \in T/N$.
We can construct ${\cal A}$ as
the algebra generated by the symbols $W([x])$
subject to the formal multiplication rule
\begin{equation}
W([x]) \, W([y]) =
\int_{N} \int_{N} W(x+n) W(y+m) \, V_{\Omega}(dn) \, V_{\Omega}(dm) \; ,
\end{equation}
where $W(x)$ are the generators of the Weyl algebra of $(T, \Omega )$,
and $x, \; y \in T$. Here the symbol $V_{\Omega}(dn)$
denotes, formally, the ``canonical" symplectic
measure on $(N, \Omega )$. More precisely, when $N$ has finite (even)
dimension $q$, this measure is given by the $q$-form
\begin{equation}
V_{\Omega} (dn) \equiv \frac{1}{(4 \pi )^{q/2}} \, \Omega \wedge \cdots
\wedge \Omega \; ,
\end{equation}
where the wedge product has $q/2$ terms. We assume, for the purpose of
our heuristic argument, that an infinite-dimensional
analogue of $V_{\Omega}(dn)$ also exists.
Clearly, the right hand side of the product Eq.\,(29)
depends only on the equivalence classes $[x], \; [y] \in T/N $.
Also, combining Eq.\,(29) with Eq.\,(11) yields
\begin{eqnarray}
W([x]) \, W([y]) & = & \int_{N} \int_{N}
e^{-\frac{i}{2} \Omega (x+n,y+m)}
\, W(x+y+n+m) \, V_{\Omega}(dn) \, V_{\Omega}(dm) \nonumber \\
& = & W([x+y]) \int_{N} \int_{N} V_{\Omega}(dn) \, V_{\Omega}(dm)
\, e^{- \frac{i}{2} \Omega (x+n,y+m)} \; ,
\end{eqnarray}
where in the second line we formally identified all $W(x+y+n)$ with
$W([x+y])$ for $n \in N$. When $T$ admits the decomposition given by
Eq.\,(25), the complex number multiplying $W([x+y])$ in Eq.\,(31)
can be evaluated easily to be
\begin{eqnarray}
& & e^{- \frac{i}{2} \Omega (x_{\perp}, y_{\perp}) }
\int_{N} \int_{N} V_{\Omega}(dn) \, V_{\Omega}(dm) \,
e^{- \frac{i}{2} \Omega (n,m) } \nonumber \\
& = & e^{- \frac{i}{2} \Omega (x_{\perp}, y_{\perp}) }
\; ,
\end{eqnarray}
where, using the measure Eq.\,(30),
the last manipulation can be justified rigorously when $N$ is finite
dimensional. Thus, it is possible to regard the expression
\begin{equation}
e^{- \frac{i}{2} \sigma ([x],[y]) } =
\int_{N} \int_{N} V_{\Omega}(dn) \, V_{\Omega}(dm) \, e^{- \frac{i}{2}
\Omega (x+n,y+m)} \;
\end{equation}
as a formal definition for the symplectic inner product $\sigma$ on the
quotient space $Q = T/N$.

We now show that the theory ${\cal A}$
just constructed coincides with the ordinary
Klein-Gordon field theory (Sect.\,1) when $(M,g)$ is globally
hyperbolic. We take $S(M)$ to be the space of all $C^{\infty}$ global
solutions of the Klein-Gordon equation (2) on $(M,g)$. Given $x \in T$,
$x = \sum_{i} (F_{i}, U_{i})$, consider the global solutions $F_{i}^{G}
\in S_{0}(M)$ defined by
\begin{equation}
F_{i}^{G} \equiv i_{U_{i} M} F_{i} \;
\end{equation}
[note that, since $(M,g)$ is globally hyperbolic, $M$ itself is a locally
causal subset]. We can write
\begin{equation}
x = \sum_{i} (F_{i}, U_{i}) = \left[ x - (\sum_{i}F_{i}^{G}, M) \right]
+
(\sum_{i}F_{i}^{G}, M) \; .
\end{equation}
It is now easy to verify that the first term in Eq.\,(35), namely $x -
(\sum_{i}F_{i}^{G} , M)$, is an element of the subset $N$ [Eq.\,(24)],
and that the second term, $(\sum_{i}F_{i}^{G}, M)$, is orthogonal (under
$\Omega$) to $N$ [this second term is equivalent, under $\Omega$ inner
product with elements of $T$, to the global solution
$\sum_{i}F_{i}^{G} \in S(M)$, so the result follows from the definition of
$N$]. This means we have, in Eq.\,(35), the orthogonal decomposition Eq.\,(25)
already performed for us. In particular, $x_{\perp} =
(\sum_{i}F_{i}^{G}, M)$. Hence,
for $x = \sum_{i} (F_{i} , U_{i})$, $y= \sum_{j} (H_{j} , V_{j}) $,
\begin{eqnarray}
\sigma ( [x],[y] ) & = & \Omega (x_{\perp}, y_{\perp} ) \nonumber \\
& = & \Omega [(\sum_{i}F_{i}^{G}, M), (\sum_{j} H_{j}^{G}, M)] \nonumber
\\
& = & {\sigma}_{M} (\sum_{i}F_{i}^{G}, \sum_{j}H_{j}^{G} ) \; ,
\end{eqnarray}
where $\sigma_{M}$ denotes the global Klein-Gordon inner product of the
spacetime $(M,g)$. Therefore we recover the ordinary symplectic space of
solutions: $(Q, \sigma ) \cong (S_{0}, {\sigma}_{M})$, and the
field theory ${\cal A}$ defined by Eqs.(28) on a globally hyperbolic
spacetime $(M,g)$ is isomorphic to the standard Klein-Gordon theory
as constructed in Sect.\,1.

The construction we just described for the field algebra of an
arbitrary spacetime is quite rigid: once the space of global solutions,
$S(M)$, is specified, the local algebra ${\cal A}^{(KG,m)}$ is
determined completely by the geometry of $(M,g)$. It is clear that for
many spacetimes the resulting field theory will not satisfy locality as
formulated in Sect.\,2. We will call a spacetime $(M,g)$ ``micro-causal"
with respect to the Klein-Gordon field if (i) the local algebra ${\cal
A}^{(KG,m)}$ defines a strongly local (Sect.\,2) quantum field theory on
$M$, and (ii) every $p \in M$ has a locally causal neighborhood $U$ such
that
\begin{equation}
U \cap C(p) \subset J^{+}(p,U) \cup J^{-}(p,U) \; .
\end{equation}
One important conclusion we can derive from the Theorem stated at the
end of Sect.\,2 is: if $(M,g)$ is micro-causal and $U$, $V \subset M$
are spacelike-separated open sets, then the subalgebras ${\cal A}(U)$
and ${\cal A}(V)$ commute. Since otherwise, according to the Theorem in
Sect.\,2, there would be points $p \in U$ and $q \in V$ such that $q \in
C(p)$, and this is in contradiction with
the spacelike separation of the open sets
$U$ and $V$ because in a micro-causal spacetime any connector is
necessarily a causal curve.
We will return to a discussion of micro-causality briefly in Sect.\,5
after studying some simple examples in the next section.

We turn now, as promised, to the construction of the symplectic form
$\sigma$ on $Q=T/N$ in the general case where the
decomposition Eq.\,(25) does not necessarily exist. It will be
instructive to first recall the following, more familiar construction: Let
$(A, <,>)$ be a (real) inner-product
space (not necessarily complete), and $P \subset A$ a closed subspace. Then,
on the quotient space $A/P$
there exists a canonical (symmetric, positive-definite)
inner product, $<,>_{\perp}$,
which can be defined as follows.
Given $x \in A$, let $\delta = \inf_{s \in P}
|| x - s || $, where $||x||$ denotes the norm $<x,x>^{1/2}$,
and, since $P$ is closed, $\delta > 0 $ unless $x \in P$.
Find a sequence $\{ x_{k} \in P
\}$ such that $||x - x_{k}|| \rightarrow \delta$. It is easy to show,
either from the polarization identity for the norm $|| \cdot ||$, or
directly, using the linearity of $P$ and the triangle inequality, that
for every $x \in A$ $\; \{ x_{k} \}$ chosen as above is a Cauchy sequence.
Now for any pair of vectors $x, \; y \in A$
define $<x , y>_{\perp} \equiv \lim_{k \rightarrow \infty}
<x - x_{k},y - y_{k}>$. It can be verified that this limit (which exists since
$\{ <x - x_{k},y - y_{k}> \}$ is a Cauchy
sequence in $\bf R$) is independent of
the choice of Cauchy sequences $\{ x_{k} \} , \; \{ y_{k} \} \subset P $.
Also, it is clear that $<x,y>_{\perp}$ thus defined depends only on
the equivalence classes $[x], \; [y] \in A/P$. Hence this
construction of $<,>_{\perp}$
gives a well-defined inner product on $A/P$.

Now let $(T, \Omega )$ be
a symplectic space. Under the family of seminorms, $\{ p_{x}
(y) \equiv | \Omega (x,y) | , \; x \in T \}$, $(T, \Omega )$ is a locally
convex topological vector space ([5], Chapter IV). In this topology,
$U$ is an open neighborhood of $0$ iff $\forall x \in U$ there exist
an $\epsilon > 0$ and finitely many $v_{i} \in T$ such that $\{ y \in T |
p_{v_{i}}(y-x) < \epsilon \} \subset U$. If $N \subset T$ is a closed
subspace [our subspace $N$ defined by Eq.\,(24) is closed],
we can define a symplectic
form $\sigma$ on $T/N$ by a construction entirely similar to that of
$<,>_{\perp}$ on $A/P$.
However, since the topology of $T$ is determined by an uncountable family of
seminorms rather than by a single norm, we need to use Cauchy nets
instead of Cauchy sequences. Recall that a net is a map from a
directed index set into $T$. A directed set is a partially ordered set
$\Lambda$ with the property that given any pair $\alpha , \; \beta \in
\Lambda$ there exists a $\gamma \in \Lambda$ such that $\gamma \geq \alpha$
and $\gamma \geq \beta$. A net $\{ x_{\alpha} \}$ is Cauchy if given any open
neighborhood $U$ of $0$ we can find a $\gamma \in \Lambda$ such that
$(x_{\alpha} - x_{\beta} ) \in U$ for
all $\alpha , \; \beta \geq \gamma$, and it converges to
$x_{0}$ if for all such $U$ we can find a $\beta \in \Lambda$ such
that $(x_{\alpha} - x_{0}) \in U$ for all $\alpha \geq \beta$.
Now fix the index set $\Lambda$ to be the directed set
consisting of all finite subsets of $T$, partially
ordered by inclusion. For each $\alpha \in \Lambda$, let $P_{\alpha}$
denote the seminorm $P_{\alpha} (x) \equiv \sum_{\alpha}
p_{v_{\alpha}} (x) $, where the sum is over the (finitely many)
vectors $v_{\alpha}$ contained in $\alpha$. Given $x \in T$, we
construct a Cauchy net $\{ x_{\alpha} \} \subset N$ as follows. Let
for each $\alpha \in \Lambda$ $\; {\delta}_{\alpha} \equiv
\inf_{n \in N} P_{\alpha}(x-n)$. Pick a vector $x_{\alpha} \in N$ such
that $| P_{\alpha}(x-x_{\alpha}) - {\delta}_{\alpha} |
< 1/| \alpha |$, where $| \alpha |$
denotes the number of elements in $\alpha \in \Lambda$. It is not
difficult to show that the net $\{ x_{\alpha} \}$ thus
constructed is Cauchy. Now, for any
pair of vectors $x, \; y \in T$, define $\sigma (x,y) \equiv
\lim_{\alpha} \Omega ( x - x_{\alpha} , y - y_{\alpha} )$. This limit
exists since the net $\{ \Omega (
x - x_{\alpha} , y - y_{\alpha} ) \}$ is Cauchy in $\bf R$ (hence converges to
a unique real number). It can also be verified that the limit
is independent of the Cauchy nets $\{ x_{\alpha} \} , \; \{ y_{\alpha}
\} \subset N$ chosen, and depends only on the equivalence classes $[x],
\; [y] \in T/N$. Hence $\sigma$ is the desired symplectic structure on
$T/N$.

{\bf \noindent 4. Some simple non-globally-hyperbolic examples}

In this section we will very briefly discuss Klein-Gordon field theory
on two examples, both flat, of
non-globally-hyperbolic two-dimensional spacetimes. Our discussion is
limited to the construction of the local field algebra on these spacetimes
in accordance with the
general ideas of Sect.\,3.

Our first example is a two-dimensional torus $T^{2}$
with a flat Lorentz metric $\eta$
and a massless Klein-Gordon field. Flat Lorentzian tori admit closed
timelike curves through every point. A detailed analysis and
classification of these Lorentz tori can be found in Sect.\,IV of
[10]. Here we will work with the simplest such tori:
two-dimensional Minkowski spacetime modulo the
translation subgroup generated by two orthogonal vectors
(generators). We assume, for
simplicity, that the ratio of the length of the spacelike generator to
the length of the orthogonal timelike one is an integer $p$. When this ratio
is irrational, the Lorentz torus does not admit any non-constant
global solutions to
the massless Klein-Gordon equation, so its field algebra would be
trivial [${\cal A} (U) = {\bf C} \; \forall \; \mbox{open} \; U \subset
M$; i.e. a completely degenerate field theory]. When $p$ is a
non-integer but rational number the resulting structure of the field algebra
is essentially the same as the integer case.
Without loss of generality, we assume that the timelike generator of
$(T^{2}, \eta )$ has length $1$ and the spacelike generator length $p$.

We take $S(M)$ to be the space of all $C^{\infty}$ global solutions of
${\nabla}^{a}{\nabla}_{a} \phi = 0$ on $(T^{2}, \eta )$. Let $p = 1$.
Given an
element $(F,U) \in T$, we construct the data $\{ F_{0},
\dot{F}_{0} \}$ that $F$ induces on some global spacelike hypersurface of
the form $\{ t = \mbox{constant} \}$. By Eq.\,(12) of
[10], global solutions $\phi$ are constrained to satisfy $\int
\dot{\phi}_{0} \, ds = 0$, where $\int ds$ denotes integration over the
global $\{ t = \mbox{constant} \}$ hypersurface.
It is not difficult to show, then, that the
symplectic inner product $\sigma$
between two equivalence classes $[(F,U)]$ and
$[(G,V)] \in T/N$ is given by
\begin{eqnarray}
\sigma ( [(F,U)] , [(G,V)] ) & = & \nonumber \\
\int [ F_{0} (\dot{G}_{0} - <\dot{G}_{0}>) & - & G_{0} (\dot{F}_{0}
- <\dot{F}_{0}>) ] \, ds \; ,
\end{eqnarray}
where $< \cdots >$ denotes the average
\begin{equation}
< f > \equiv \int f \, ds \; .
\end{equation}
For $p \geq 2$,
let $\pi$ denote the finite group of
order $p$ consisting of the discrete isometries generated by
a unit translation in the $x$-direction,
and let for any $f \in C^{\infty} (T^{2}) $
$\; \pi f$ denote the function obtained by summing all
$p$ images of $f$ under
the action of this group.
Then the symplectic
product $\sigma$ on $T/N$ is
\begin{eqnarray}
\sigma ( [(F,U)] , [(G,V)] ) & = & \nonumber \\
\frac{1}{p} \int [ \pi F_{0} (\pi \dot{G}_{0} - <\pi \dot{G}_{0}>) & -
& \pi
G_{0} (\pi \dot{F}_{0}
- <\pi \dot{F}_{0}>) ] \, ds \; ,
\end{eqnarray}
where
\begin{equation}
< f > \equiv \frac{1}{p} \int f \, ds \; .
\end{equation}

As they stand, the field theories defined by Eqs.\,(38)$-$(41)
do not satisfy locality. However, as is well known,
in two dimensions
the presence of infrared divergences in the two-point
function makes it necessary
to smear a massless scalar field only
with those
test functions $f \in C_{0}^{\infty} (M)$ that satisfy $\int_{M} f \,
d \tau = 0$. This is easily seen to be equivalent to building
the space $T$ so that all local solutions $F = {\cal E}_{U} f$
satisfy $<\dot{F}_{0}> = 0$. With this choice of $T$, the
symplectic product $\sigma$ is exactly the same as it would be for the
globally hyperbolic cylinder $S^{1} \times {\bf R}$, obtained by undoing
the identification in the time direction of $(T^{2} , \eta )$. Therefore,
after this modification,
the massless Klein-Gordon theory on $(T^{2}, \eta )$ does satisfy strong
locality as formulated in Sect.\,2. In fact, in both the cases $p=1$ and $p
\geq 2$, and for any $q \in T^{2}$, the union of the two closed
null geodesics passing through $q$ constitute the set $C(q)$. Thus
micro-causality (Sect.\,3) holds for these examples.

Our next example is the two-dimensional spacetime (with timelike
singularities) consisting of the open strip
$\{ 0 < x < 1 , \; -\infty < t <
\infty \}$ in
Minkowski space (the singularities are at the walls $\{ x=0 \}$ and $\{
x = 1 \}$). We let $S(M)$ be the space of all
solutions of the (massless or massive) Klein-Gordon equation which are
$C^{\infty}$ on $M$ and
$C^{1}$ at the boundaries $\{ x=0 \}$ and $\{ x=1 \}$, i.e. we constrain
the solutions in $S(M)$ to admit
$C^{1}$ extensions (not necessarily as solutions) across these
boundaries. [It is possible, but tedious, to show that the field algebra
remains exactly the same even when this constraint on the boundary
behavior of $S(M)$ is relaxed completely.] Let for each $(F,U) \in T$
$\; F^{B}$ denote the unique global solution obtained by the following
construction: Consider a Cauchy surface $\Sigma$
for $(U, {\eta}|_{U})$ and
extend it to a global spacelike hypersurface reaching from $x=0$ to
$x=1$. The local solution $F$ induces well-posed initial data on this
surface which vanish outside of $\Sigma$. Now let $F^{B}$ be the unique
solution of the Klein-Gordon equation $({\nabla}^{a}{\nabla}_{a} - m^{2})
\phi = 0 $ that evolves from these data with
the boundary conditions $\phi = 0$ on $\{ x=0 \}$ and
$\phi = 0$ on $\{ x=1 \}$.
Then, it is not difficult to show that the symplectic inner product
$\sigma$ between two equivalence classes $[(F,U)], \; [(G,V)] \in T/N$ is
given by
\begin{equation}
\sigma ( [(F,U)], [(G,V)] ) =
\int (F_{0}^{B} \dot{G}_{0}^{B} - G_{0}^{B} \dot{F}_{0}^{B} ) \, ds \; ,
\end{equation}
where the integral is over any global spacelike hypersurface of the form
$\{ t = \mbox{constant} \}$ (independence of the surface is insured by
the boundary conditions on $F^{B}$ and $G^{B}$). With $T/N$ endowed
with this symplectic
structure, the field theory ${\cal A}^{(KG,m)}$ defined by
Eqs.\,(28) satisfies micro-causality on $(M, \eta )$ for all values of $m$.

{\bf \noindent 5. Conclusions}

With the general construction we presented in Sect.\,3, it now
becomes possible to discuss quantum field theory on spacetimes with naked
singularities or closed timelike curves in the clean framework of the
algebraic approach. For spacetimes with naked singularities, our
construction does not eliminate the issue of what boundary conditions
are to be imposed on the fields ``at" the singularities, but rather
delegates this problem to the description of the states of the quantum
field instead of the fields themselves. More precisely,
although our construction
gives an unambiguous description of the local field algebra throughout
spacetime,
when global hyperbolicity is violated it is no longer
sufficient to specify a
quantum state only in the vicinity of an initial spacelike hypersurface.
Rather, the action of the state (as a linear functional on local smeared
field
operators) needs to be
specified globally on the entire spacetime, a process which in general
will involve input that can be interpreted as ``boundary conditions"
(for the state) at the singularities. For spacetimes with closed
timelike curves, our construction naturally leads to a new, microscopic
causality condition, ``micro-causality," that needs to be satisfied for
a consistently local (Klein-Gordon) quantum field theory to exist.

The notion of micro-causality is quite distinct from the classical
notion of ``benignness" introduced earlier in Ref.\,[10]. Recall ([10])
that, in the present context, a spacetime $(M, g)$
is called benign if every point
has an open neighborhood $U$ such that any solution of the
Klein-Gordon equation in $U$ which is locally extendible off $U$ (i.e.
which does not develop singularities at the boundary of $U$) is
extendible to a global solution on $(M,g)$. In a benign spacetime, observers
making local
(classical) measurements of the Klein-Gordon field will be unable to
distinguish their spacetime from one which is
globally hyperbolic. The spacetimes we discussed in the
examples of Sect.\,4 are benign with respect to the Klein-Gordon
equation (massless for the first example).
Benignness and micro-causality are independent properties; i.e.
neither of them implies nor is implied by
the other. A straightforward quantum-field-theory
analogue to the notion of benignness would be the property that (i)
micro-causality holds and (ii) every
point $p \in M$ has a locally causal neighborhood $U$ with
\begin{equation}
( \left. {\cal A}^{(KG,m)} \right|_{U} , U) \cong ( \left. {\cal A}^{(KG, m)}
\right|_{[(U, g {|}_{U})]} \, , \, U) \; ,
\end{equation}
where $\cong$ denotes isomorphism in the quantum-field-theory
sense as we discussed in
Sect.\,2 above, and the field theory on the right hand side denotes the
standard Klein-Gordon theory for the globally hyperbolic spacetime $(U,
g|_{U})$. Thus, spacetime is quantum-benign if
and only if both strong locality (condition SL)
and condition (43) are satisfied.
Local measurements of the quantum field operator (and
other observables derived from it) in a spacetime satisfying (43)
will be inadequate to distinguish that spacetime from one
which is globally hyperbolic. The
spacetimes discussed in the examples in Sect.\,4 are benign also in this
quantum sense.

As long as attention is focused on the standard Klein-Gordon or Maxwell
field theories, rather than on some generalized version of these where
the space of classical solutions is suitably restricted, the
micro-causality requirement on
spacetimes with closed timelike curves is a much more severe
constraint than the classical requirement of
benignness. Indeed, it is unlikely, for example, that there exists {\it
any} four-dimensional compact spacetime which satisfies micro-causality
and/or the condition (43) with respect to the Klein-Gordon field (see
[10] for a discussion of classical field theory on compact spacetimes).
Uncovering the precise status of micro-causality in spacetimes
with closed timelike curves is of obvious importance for understanding
the physics of causality violation, and this issue is going to be
dealt with in more detail in a forthcoming paper ([11]).

\newpage

{\bf \noindent REFERENCES}

\noindent{\bf 1.} J. Dimock, Commun. Math. Phys. {\bf 77}, 219 (1980);
F. G. Friedlander, {\it The Wave Equation in Curved Spacetime}
(Cambridge University Press, Cambridge, 1973).

\noindent{\bf 2.} B. S. Kay and R. M. Wald, Phys. Reports {\bf 207}, 49
(1991).

\noindent{\bf 3.} R. M. Wald, in: {\it Quantum Mechanics in Curved
Spacetime}, J. Audretsch and V. de Sabbata eds. (Plenum, New York,
1992).

\noindent{\bf 4.} B. Simon, in: {\it Mathematics of Contemporary
Physics}, R. F. Streater ed. (Academic Press, New York, 1972).

\noindent{\bf 5.} J. H. Conway, {\it A Course in Functional Analysis}
(Springer Verlag, New York, 1990).

\noindent{\bf 6.} A. Arveson, {\it An Invitation to $C^{\ast}$ Algebras}
(Springer Verlag, New York, 1972).

\noindent{\bf 7.} U. Yurtsever, {\it The origin of spacetime topology
and generalizations of quantum field theory}, UCSB Physics Preprint
UCSBTH-92-45, (October 1992).

\noindent{\bf 8.} S. W. Hawking and G. F. R. Ellis, {\it The large Scale
Structure of Spacetime} (Cambridge University Press, Cambridge, 1973).

\noindent{\bf 9.} J. Friedman, G. Klinkhammer, F. Echeverria, M. S.
Morris, I. D. Novikov, K. S. Thorne, and U. Yurtsever, Phys. Rev. D {\bf
42}, 1915 (1990).

\noindent{\bf 10.} U. Yurtsever, J. Math. Phys. {\bf 31}, 3064 (1990).

\noindent{\bf 11.} U. Yurtsever, {\it Quantum theory, locality, and
closed timelike curves}, UCSB Physics Preprint (in preparation).

\end{document}